\documentclass[twocolumn]{aastex631}

\usepackage{amsmath}
\usepackage{array}   
\usepackage{enumitem}
\usepackage{CJK}
\usepackage{hyperref}
\received{}
\revised{}
\accepted{}
\submitjournal{\apjl}

\begin{document}

\title{GOALS-JWST: Revealing the Buried Star Clusters in the Luminous Infrared Galaxy VV 114}

\correspondingauthor{S. T. Linden}
\email{slinden@umass.edu}

\author[0000-0002-1000-6081]{Sean T. Linden}
\affiliation{Department of Astronomy, University of Massachusetts at Amherst, Amherst, MA 01003, USA}

\author[0000-0003-2638-1334]{Aaron S. Evans}
\affil{Department of Astronomy, University of Virginia, 530 McCormick Road, Charlottesville, VA 22903, USA}
\affiliation{National Radio Astronomy Observatory, 520 Edgemont Road, Charlottesville, VA 22903, USA}

\author[0000-0003-3498-2973]{Lee Armus}
\affiliation{IPAC, California Institute of Technology, 1200 E. California Blvd., Pasadena, CA 91125}

\author[0000-0002-5807-5078]{Jeffrey A. Rich}
\affiliation{The Observatories of the Carnegie Institution for Science, 813 Santa Barbara Street, Pasadena, CA 91101}

\author[0000-0003-3917-6460]{Kirsten L. Larson}
\affiliation{AURA for the European Space Agency (ESA), Space Telescope Science Institute, 3700 San Martin Drive, Baltimore, MD 21218, USA}

\author[0000-0001-8490-6632]{Thomas Lai}
\affiliation{IPAC, California Institute of Technology, 1200 E. California Blvd., Pasadena, CA 91125}

\author[0000-0003-3474-1125]{George C. Privon}
\affiliation{National Radio Astronomy Observatory, 520 Edgemont Road, Charlottesville, VA 22903, USA}
\affiliation{Department of Astronomy, University of Florida, P.O. Box 112055, Gainesville, FL 32611, USA}

\author[0000-0002-1912-0024]{Vivian U}
\affiliation{Department of Physics and Astronomy, 4129 Frederick Reines Hall, University of California, Irvine, CA 92697, USA}

\author[0000-0003-4268-0393]{Hanae Inami}
\affiliation{Hiroshima Astrophysical Science Center, Hiroshima University, 1-3-1 Kagamiyama, Higashi-Hiroshima, Hiroshima 739-8526, Japan}

\author[0000-0002-4375-254X]{Thomas Bohn}
\affil{Hiroshima Astrophysical Science Center, Hiroshima University, 1-3-1 Kagamiyama, Higashi-Hiroshima, Hiroshima 739-8526, Japan}

\author[0000-0002-3139-3041]{Yiqing Song}
\affiliation{Department of Astronomy, University of Virginia, 530 McCormick Road, Charlottesville, VA 22903, USA}
\affiliation{National Radio Astronomy Observatory, 520 Edgemont Road, Charlottesville, VA 22903, USA}

\author[0000-0003-0057-8892]{Loreto Barcos-Mu\~noz}
\affiliation{National Radio Astronomy Observatory, 520 Edgemont Road, Charlottesville, VA 22903, USA}
\affiliation{Department of Astronomy, University of Virginia, 530 McCormick Road, Charlottesville, VA 22903, USA}

\author[0000-0002-2688-1956]{Vassilis Charmandaris}
\affiliation{Department of Physics, University of Crete, Heraklion, 71003, Greece}
\affiliation{Institute of Astrophysics, Foundation for Research and Technology-Hellas (FORTH), Heraklion, 70013, Greece}
\affiliation{School of Sciences, European University Cyprus, Diogenes street, Engomi, 1516 Nicosia, Cyprus}

\author[0000-0001-7421-2944]{Anne M. Medling}
\affiliation{Department of Physics \& Astronomy and Ritter Astrophysical Research Center, University of Toledo, Toledo, OH 43606, USA}
\affiliation{ARC Centre of Excellence for All Sky Astrophysics in 3 Dimensions (ASTRO 3D)}

\author[0000-0002-2596-8531]{Sabrina Stierwalt}
\affiliation{Occidental College, Physics Department, 1600 Campus Road, Los Angeles, CA 90042}

\author[0000-0003-0699-6083]{Tanio Diaz-Santos}
\affiliation{Institute of Astrophysics, Foundation for Research and Technology-Hellas (FORTH), Heraklion, 70013, Greece}
\affiliation{School of Sciences, European University Cyprus, Diogenes street, Engomi, 1516 Nicosia, Cyprus}

\author[0000-0002-5666-7782]{Torsten B\"oker}
\affiliation{European Space Agency, Space Telescope Science Institute, Baltimore, Maryland, USA}

\author[0000-0001-5434-5942]{Paul van der Werf}
\affiliation{Leiden Observatory, Leiden University, PO Box 9513, 2300 RA Leiden, The Netherlands}

\author[0000-0002-5828-7660]{Susanne Aalto}
\affiliation{Department of Space, Earth and Environment, Chalmers University of Technology, 412 96 Gothenburg, Sweden}

\author[0000-0002-7607-8766]{Philip Appleton}
\affiliation{IPAC, California Institute of Technology, 1200 E. California Blvd., Pasadena, CA 91125}

\author[0000-0002-1207-9137]{Michael J. I. Brown}
\affiliation{School of Physics \& Astronomy, Monash University, Clayton, VIC 3800, Australia}

\author[0000-0003-4073-3236]{Christopher C. Hayward}
\affiliation{Center for Computational Astrophysics, Flatiron Institute, 162 Fifth Avenue, New York, NY 10010, USA}

\author[0000-0001-6028-8059]{Justin H. Howell}
\affiliation{IPAC, California Institute of Technology, 1200 E. California Blvd., Pasadena, CA 91125}

\author[0000-0002-4923-3281]{Kazushi Iwasawa}
\affiliation{Institut de Ci\`encies del Cosmos (ICCUB), Universitat de Barcelona (IEEC-UB), Mart\'i i Franqu\`es, 1, 08028 Barcelona, Spain}
\affiliation{ICREA, Pg. Llu\'is Companys 23, 08010 Barcelona, Spain}

\author[0000-0003-2743-8240]{Francisca Kemper}
\affiliation{Institut de Ciencies de l\'Espai (ICE, CSIC), Can Magrans, s/n, 08193 Bellaterra, Barcelona, Spain}
\affiliation{ICREA, Pg. Lluis Companys 23, Barcelona, Spain}
\affiliation{Institut d\'Estudis Espacials de Catalunya (IEEC), E-08034 Barcelona, Spain}

\author[0000-0003-1924-1122]{David T. Frayer}
\affiliation{Green Bank Observatory, 155 Observatory Road, Green Bank, WV 24944, USA}

\author[0000-0002-9402-186X]{David Law}
\affiliation{Space Telescope Science Institute, 3700 San Martin Drive, Baltimore, MD, 21218, USA}

\author[0000-0001-6919-1237]{Matthew A. Malkan}
\affiliation{Department of Physics \& Astronomy, UCLA, Los Angeles, CA 90095-1547}

\author[0000-0001-7712-8465]{Jason Marshall}
\affiliation{Glendale Community College, 1500 N. Verdugo Rd., Glendale, CA 91208}

\author[0000-0002-8204-8619]{Joseph M. Mazzarella}
\affiliation{IPAC, California Institute of Technology, 1200 E. California Blvd., Pasadena, CA 91125}

\author[0000-0001-7089-7325]{Eric J. Murphy}
\affiliation{National Radio Astronomy Observatory, 520 Edgemont Road, Charlottesville, VA 22903, USA}

\author[0000-0002-1233-9998]{David Sanders}
\affiliation{Institute for Astronomy, University of Hawaii, 2680 Woodlawn Drive, Honolulu, HI 96822}

\author[0000-0001-7291-0087]{Jason Surace}
\affiliation{IPAC, California Institute of Technology, 1200 E. California Blvd., Pasadena, CA 91125}

\begin{abstract}
We present the results of a {\it James Webb Space Telescope} NIRCam investigation into the young massive star cluster (YMC) population in the luminous infrared galaxy VV 114. We identify 374 compact YMC candidates with a $S/N \geq 3$, 5, and 5 at F150W, F200W, and F356W respectively. A direct comparison with our {\it HST} cluster catalog reveals that $\sim 20\%$ of these sources are undetected at optical wavelengths. Based on {\it yggdrasil} stellar population models, we identify 17 YMC candidates in our {\it JWST} imaging alone with F150W-F200W and F200W-F356W colors suggesting they are all very young, dusty ($A_{V} = 5 - 15$), and massive ($10^{5.8} < M_{\odot} < 10^{6.1}$). The discovery of these `hidden' sources, many of which are found in the `overlap' region between the two nuclei, quadruples the number of $t < 3$ Myr clusters, and nearly doubles the number of $t < 6$ Myr clusters detected in VV 114. Now extending the cluster age distribution ($dN/d\tau \propto \tau^{\gamma}$) to the youngest ages, we find a slope of $\gamma = -1.30 \pm 0.39$ for $10^{6} < \tau  (\mathrm{yr}) < 10^{7}$, which is consistent with the previously determined value from $10^{7} < \tau  (\mathrm{yr}) < 10^{8.5}$, and confirms that VV 114 has a steep age distribution slope for all massive star clusters across the entire range of cluster ages observed. Finally, the consistency between our {\it JWST}- and {\it HST}-derived age distribution slopes indicates that the balance between cluster formation and destruction has not been significantly altered in VV 114 over the last 0.5 Gyr.

\end{abstract}


\section{Introduction}

Studies of star formation in nearby galaxies across a wide variety of environments reveal that the majority of star formation occurs in stellar clusters, groups, and associations \citep[e.g.,][]{ll03}. Star formation activity is particularly intense within starburst and merging galaxies, where thousands of compact clusters ($R_\mathrm{eff} \sim$3-5 pc), many with masses $M \geq 10^{4} M_{\odot}$ are readily observed \citep[e.g.,][]{bcm10,mulia15,rand19,aa20a}. At the earliest phases of their evolution, these young massive star clusters (YMCs) will experience strong feedback from O-star winds, and may become unbound as the remaining gas in the parent giant molecular cloud (GMC) is expelled within a crossing time \citep{krause20}.~However, extremely high-pressure regions of the interstellar medium (ISM) will increase the difficulty in dispersing gas from the cluster, and may actually result in a higher cluster formation efficiency \citep[CFE;][]{dk11}.

Observations of YMCs in nearby galaxies provide evidence for fundamental differences in how feedback operates in different environments. For example, \citet{bcm02b} identified a small population of `very red clusters' in the Antennae merger whose optical colors suggest that they are young ($t = 4 - 6$ Myr), massive ($M = 10^{5-6} M_{\odot}$), and very dusty ($A_{V} = 5 - 7$). In contrast, observations of the low-metallicity dwarf NGC 4449 reveal that many of the youngest clusters ($M \sim 10^{4} M_{\odot}$, $t = 3 - 5$ Myr) have relatively low dust extinction ($A_{V} \leq 1.5$), suggesting an efficient removal of the remaining gas and dust \citep{reines08b}.

Simulations suggest that YMCs formed within the densest regions of the ISM may host ionized stellar winds that slow down due to critical radiative cooling, ultimately allowing the cluster to retain fuel for ongoing (or future) generations of star formation \citep{silich20}. Constraining this short-lived phase in YMC evolution remains a critical missing piece for understanding how star cluster formation proceeds in different environments.

Before the onset of stellar feedback, YMCs are deeply embedded within their birth clouds of gas and dust and suffer large internal extinction ($A_{V} \sim 7-10$), rendering them effectively un-observable at UV-optical wavelengths. Therefore, these objects are consistently missed in studies of early cluster feedback and evolution that rely on UV-optical based diagnostic tools \citep{hollyhead16,hannon22}. At longer wavelengths, near- and mid-infrared emission can be used to trace the dust cocoons surrounding newly-formed YMCs \citep{corbelli17,mm21}. However, the resolution of {\it Spitzer} at $\lambda = 3 - 5\mu$m ($\sim 2"$) only allowed individual clusters to be resolved for galaxies at very close distances, making it difficult to build meaningful samples of embedded YMCs at higher-redshifts and in different environments.

Luminous IR galaxies (LIRGs: defined as $L_{\rm IR} [8-1000\mu {\rm m}] \geq 10^{11.0}$ L$_\odot$) host the most extreme stellar nurseries in the local Universe. The activity in LIRGs is primarily triggered by gas-rich galaxy interactions, and at their peak, local LIRGs have star formation rates (SFRs) $\sim 100$ times higher than the Milky Way, placing them well above the star formation main sequence (SFMS) \citep{speagle14}. Many local LIRGs have individual star-forming clumps with sizes of $50 - 100$pc and SFR surface densities ($\Sigma_{SFR}$) of $0.1 - 10$ $M_{\odot}$yr$^{-1}$kpc$^{-2}$) that rival those of high-$z$ galaxies \citep{klarson20}. Therefore, local LIRGs represent the ideal laboratory for studying YMC formation and evolution in extreme environments at high-resolution.

With the unprecedented sensitivity and resolving power of the {\it James Webb Space Telescope} ({\it JWST}) we are now able to identify and characterize individual YMCs forming in the densest and dustiest regions of the ISM in galaxies out to $\sim 100$ Mpc in the near-infrared (NIR). In this letter we present {\it JWST} NIR observations of the embedded YMC population in the starburst LIRG VV 114 (VV 114 = IC 1623). VV 114 is classified as an early-stage merger with the two galaxies aligned East-West and a projected nuclear separation of $\sim 8$ kpc (VV 114E and VV 114W respectively). Previous {\it HST} observations of the galaxy have been used to characterize the UV-bright population of clusters found almost exclusively in VV 114W \citep{linden21}. The observations described in this paper represent the most comprehensive census to date of YMC formation and evolution in VV 114 in the NIR, and sets the stage for future {\it JWST} studies of YMC formation in nearby LIRGs. 

Throughout this letter, we adopt a WMAP Cosmology of $H_0 = 69.3$ km s$^{-1}$ Mpc$^{-1}$, $\Omega _{\rm matter} = 0.286$, and $\Omega _{\Lambda} = 0.714$ \citep[e.g., see][]{wmap}.

\begin{figure*}[]
\centerline{\includegraphics[width=4.65truein]{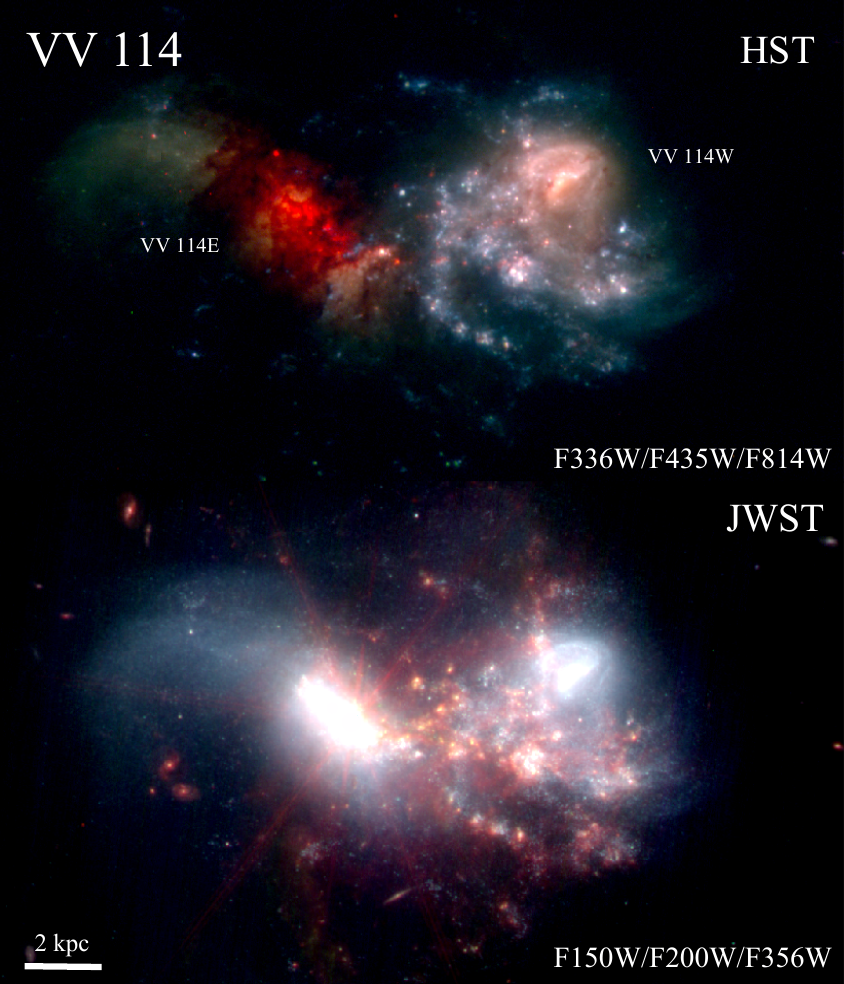}}
\caption{{\bf Top:} False-color {\it HST} imaging of VV 114 using F336W ($\lambda = 0.336 \mu$m), F435W ($\lambda = 0.435 \mu$m), and F814W ($\lambda = 0.814 \mu$m). The images are oriented N up E left, and the 2 kpc scale bar is equivalent to $4.8"$. Bright UV-optically visible clusters are found throughout VV 114W. These clusters are identified and subsequently characterized in \citet{linden21}. {\bf Bottom:} False-color {\it JWST} NIRCam image of VV 114 using F150W, F200W, and F356W observations. The {\it JWST} data reveal many bright, massive, red sources previously hidden behind the dust lane in the `overlap' region of the galaxy. The combination of the {\it JWST} filters used in this analysis allow us to characterize the properties of these red sources for the first time.}
\end{figure*}

\section{Observations}

\begin{figure*}[]
\centerline{\includegraphics[width=6.8truein]{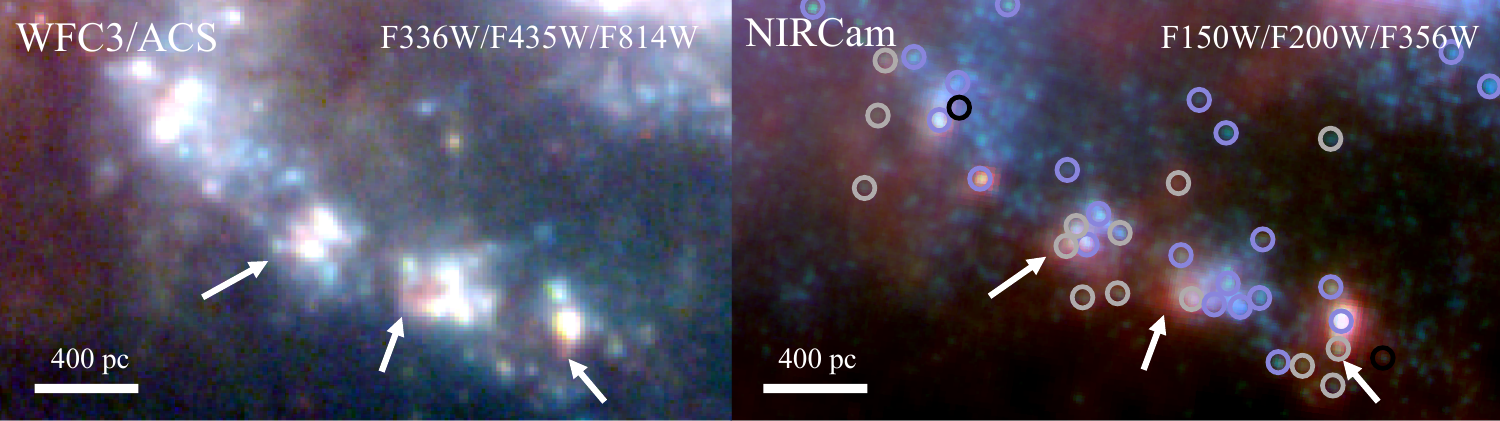}}
\caption{{\bf Left:} A false-color {\it HST} zoom-in view of the spiral arm extending from the western nucleus. {\bf Right:} A {\it JWST} view of the same spiral arm regions. Here we show the NIRCam images at native-resolution to best highlight red sources and diffuse emission surrounding star-forming regions. The sources that are extremely faint and appear to be red at UV-optical wavelengths are now seen clearly in the NIR, and are identified with white arrows. Further, we see that many of these massive star-forming regions contain a mix of clusters with very red and blue NIR colors. Compact sources identified in our {\it JWST} NIRCam images are circled in grey, sources identified in our {\it HST} images alone are circled in black, and sources which appear in both catalogs are circled in purple. Overall, by combining both {\it HST} and {\it JWST} observations, we are sensitive to the entire cycle of YMC formation in VV 114.}
\end{figure*}

{\it JWST} imaging of VV 114 was obtained as part of the Early Release Science (ERS) program ``A {\it JWST} Study of the Starburst-AGN Connection in Merging LIRGs" (PID 1328; Co-PIs: Armus, Evans). Additional studies focusing on the medium-resolution spectrometer (MRS) and mid-infrared instrument (MIRI) observations of VV 114 are presented in Rich et al. (2022 in prep) and \citet{evans22} respectively.

Near-Infrared Camera (NIRCam; \citet{nircam}) observations of VV 114 were taken on July 29th, 2022, and retrieved from the  Mikulski Archive for Space Telescopes (MAST). The galaxy was imaged using the F150W ($\lambda=1.5\mu$m) and F200W ($\lambda=2.0\mu$m) short wavelength (SW) and F356W ($\lambda=3.5\mu$m) and F444W ($\lambda=4.4\mu$m) long wavelength (LW) filters for 2319s and 580s, respectively with module B. These observations utilized the `FULL' array with the `RAPID' readout mode and the `INTRAMODEL' 3-pt dither pattern. 

The raw data have been reduced using the {\it JWST} calwebb pipeline v1.5.3 (CRDS 11.16.3). After in-flight calibration \citep{jwstcomm}, updated reference files were included in the reduction pipeline (CRDS0942). In particular, this update takes into account the higher sensitivity of NIRCam in the LW filters. To account for additional changes to the zero-point calibration (CRDS0989), we use the PHOTMJSR corrections published for module B in \citet{zpupdate2}, which are derived from a resolved stellar population analysis of M92 (PID 1334). In our analysis we focus on the two SW filters as well as the F356W to constrain the spectral energy distribution (SEDs) of individual star clusters in the NIR. These filters are the least affected by potential contributions from either a hot dust component or circumstellar dust from AGB stars, which can have a significant affect on the integrated flux of star clusters at $\lambda \geq 4 \mu$m \citep{conroy15}, and are not included in the models used to interpret the derived cluster colors (see Section 3.2)

The level-3 products (.i2d) were then aligned to the Gaia DR2 reference frame using the Drizzlepac module TweakReg \citep{gaiadr2}. Finally, our F150W, F200W, and F356W science frames were drizzled to a common scale of 0.05"/pixel, which was chosen to match the angular resolution of existing {\it HST} WFC3 and ACS UV-optical observations of VV 114, and results in a physical resolution of $\sim 20$ pc at the distance of VV 114 (84.4 Mpc). We note that this resolution is several times larger than the typical size of YMCs observed in nearby galaxies, however it is still well below the average size of cluster complexes and massive OB associations \citep[$R_{eff} \sim 100-200$ pc][]{nb06}. We further discuss the effects of resolution when interpreting our results in Section 4.

In Figures 1 and 2 we compare {\it HST} optical and {\it JWST} NIR false-color images of VV 114. At optical wavelengths we see that VV 114E and the `overlap' region between the east and west nuclei are enshrouded by dust lanes, and that detections of bright point sources at wavelengths $\lambda < 0.4\mu$m in this region are extremely limited. In contrast, the western nucleus has significantly less dust obscuration, allowing for bright blue point-sources to be easily identified throughout the disk. From the NIRCam images it is clear that at longer wavelengths our observations are sensitive to many bright, red sources, several of which were previously hidden behind the dust lane in the `overlap' region of the merger.

\section{Results}

\subsection{Cluster Identification and Photometry}

Star cluster candidates in all three NIRCam filters were selected using the program Source Extractor \citep{sex}. We considered all sources with local S/N thresholds of $\geq 3, 5,$ and 5 in 5 continuous pixels at F150W, F200W, and F356W respectively, utilizing 64 de-blending sub-thresholds. We increased the detection threshold for the F200W and F356W filters due to the increased background levels and the crowding of faint blue sources between bright star-forming regions at these wavelengths (Right Panel Figure 2). This results in 1837, 1689, and 1860 detections for each filter respectively. Finally, requiring sources to be detected in all filters, and spatially-matched to within 1 pixel, results in a final candidate list of 511 sources.

We fit 2-D elliptical Gaussian profiles to all 511 candidates in order to measure the major- and minor-axis FWHM at F150W. We then remove from our catalog sources with a major-axis FWHM that is $\geq 2\sigma_{FWHM,maj}$ larger than the FWHM of the F150W instrumental point-spread function (1.628 pixels $= 0.05"$) determined from WebbPSF \citep{perrin14}.~This requirement removes 137 candidates, including 63 objects identified by-eye as background galaxies with extended structure, and 74 additional sources that are likely larger clumps and complexes of clusters blended to appear as single objects. Here we focus on identifying compact star clusters, which can be best-modeled as single stellar populations (SSPs), and the derived properties can be compared directly to catalogs where strict size cuts have been applied. A full accounting of clumps and larger star-forming regions, which are likely a combination of multiple single stellar populations, is outside the scope of this letter.

In order to understand the overlap between the cluster catalog presented in \citet{linden21}, which was produced using {\it HST} UVIS F336W, and ACS F435W, and F814W observations, we cross reference the 179 {\it HST} sources with the 374 compact cluster candidates identified with {\it JWST}. We find that $75\%$ ($135/179$) of the sources identified with {\it HST} overlap with our NIRCam catalog within 2 pixels ($0.1"$). For the 25\% of the {\it HST} sources that we are not selecting with our {\it JWST} detection thresholds we find that several of these sources have an {\it JWST}/F200W S/N $\sim 1 - 3$ and FWHM slightly larger than the {\it JWST}/F150W PSF. Importantly, the majority of these clusters were determined to have intermediate-ages ($t \geq 20$ Myr), and thus our selection method is not biased against detecting the youngest ($t < 5$ Myr) clusters in VV 114. To push the detection thresholds lower such that we can detect all faint sources with {\it JWST} will require a detailed analysis of the completeness as a function of cluster location within the galaxy, age, mass, and size. These issues will be addressed in a follow-up paper including NIRCam imaging from all ERS targets in PID 1328. 


On the other hand, when we compare the cluster candidates that were originally identified in the {\it HST}/F814W image (4060 sources with a S/N $\geq 3$ in 5 pixels), we find that $298/374$ ($80\%$) of the compact cluster candidates identified with NIRCam were detected with {\it HST}/F814W. This fraction decreases to 60\% and 40\% for {\it HST}/F435W and {\it HST}/F336W, respectively. This comparison demonstrates that at least $20-60\%$ of the sources identified in the NIR are completely missed at optical and UV wavelengths, where clusters were identified down to limit of $m_{F336W} (\mathrm{Vega}) \sim 26$ mag \citep{linden21}.

\begin{figure*}[]
\centerline{\includegraphics[width=7.0truein]{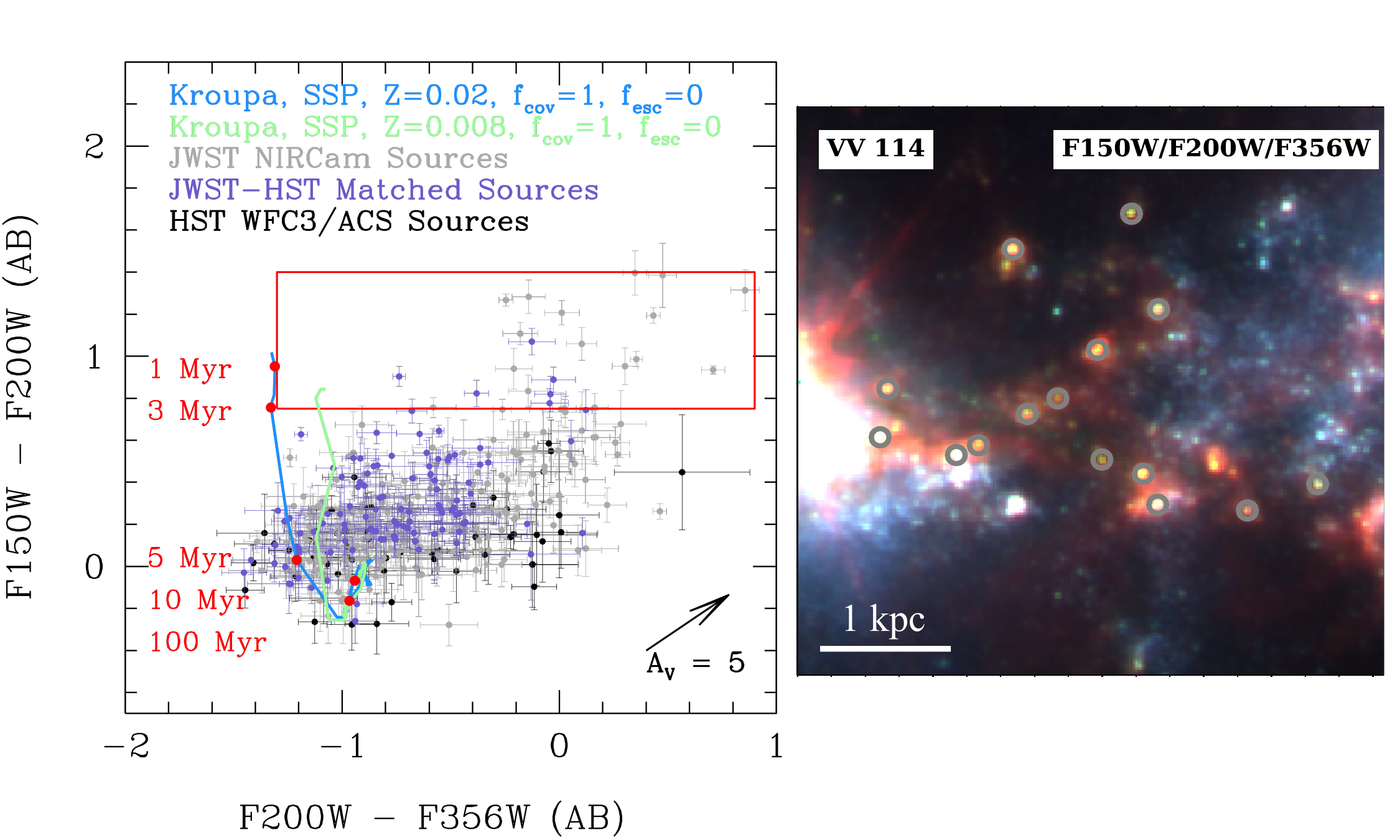}}
\caption{{\bf Left:} The F150W-F200W vs. F200W-F356W color-color diagram for all 374 compact point-sources identified in our {\it JWST} NIRCam images (grey), as well as the {\it HST} star cluster catalog presented in \citet{linden21} - black. Sources which appear in both catalogs are shown in purple. In the lower-right we show an $A_{V} = 5$ extinction vector, derived by adopting a NIR power-law slope ($\beta$) of -1.6. Overlaid in light blue and green are the {\it yggdrasil} SSP model tracks adopted here with $z=0.008$ and 0.02 as well as a $100\%$ contribution of the available ionizing photons. Ages are marked in red along the sequence. The red box highlights the region where nearly all sources are identified in the {\it JWST} imaging alone. Based on their location relative to the SSP models, these sources all appear to be very young, red, and heavily dust-enshrouded. Further, clusters with F200W-F356W $> -0.4$ and F150W-F200W $\sim 0.4$ are predominately identified with {\it JWST} NIRCam, and may represent a more evolved, yet still dust-enshrouded population of YMCs in VV 114. {\bf Right:} A zoom-in of the central region of VV 114 showing the locations of 15 of the reddest YMCs that are identified with NIRCam imaging alone (grey points). Notably, these sources lie in the `overlap' region of the merger between the east and west nuclei.}
\end{figure*}

In Figure 2 we zoom-in on the string of massive star-forming regions in the spiral arm extending from the western galaxy, that contains several UV- and optically-bright clusters (identified as the bright white sources - left Panel), as well as sources which appear very red, possibly still embedded in their birth clouds (white arrows). In the right Panel it is clear that these faint red sources are much brighter in the longer-wavelength {\it JWST} images, further suggesting that are highly obscured, and possibly young. Compact sources identified in our {\it JWST} NIRCam images are circled in grey, sources identified in our {\it HST} images alone are circled in black, and sources which appear in both catalogs are circled in purple. We note that the handful of sources which are only detected in {\it HST} images are all faint blue sources embedded in regions of diffuse emission, and are therefore not retained due to our higher S/N threshold ($\geq 5$) in the NIRCam LW filters. Overall, the right panel of Figure 2 demonstrates that many massive star-forming regions in VV 114 contain a mix of very young (red) and intermediate-age (blue) stellar clusters, many of which are detected with {\it HST}. This mixture is reminiscent of one of the most well-studied star-forming regions, 30 Doradus in the Large Magellanic Cloud, which contains both a UV-bright YMC, R136, as well as embedded YSOs in the NW region of the complex \citep{30dor}.

Photometry for 374 compact clusters identified across the three NIRCam filters was then calculated using the IDL package APER. We used an aperture of radius 2 pixels ($0.1"$), with an annulus from 4-6 pixels to measure the 3$\sigma$-clipped mean local background surrounding each cluster. Errors are estimated by varying the inner-radius of the sky annulus from 2-5 pixels. Aperture corrections for F150W, F200W, and F356W were calculated based on the encircled energy values presented in \citet{perrin14} using the WebbPSF models. We additionally applied a correction for foreground Galactic extinction, using the \citet{schlafly11} dust model and the empirical reddening law of \citet{fitz99}. We present our measurements using the AB magnitude photometric system. Finally, we repeat this procedure for the 179 {\it HST}-identified clusters in \citet{linden21} so that direct comparisons of the two catalogs can be made.

\subsection{Stellar Population Synthesis in the NIR}

For all confirmed clusters, the measured 3-band magnitudes are compared against SSP evolutionary models generated using the isochrone synthesis code {\it yggdrasil} \citep{yggdrasil}. This code computes the evolution of an instantaneous SF burst based on a Kroupa IMF and the Padova-AGB stellar evolution tracks over an age range of 1 Myr to 10 Gyr \citep[e.g.,][]{padova94}. Additionally, these models use CLOUDY \citep{cloudy17} to add the contribution from nebular emission lines by varying the fraction of ionizing photons which ionize the surrounding gas ($f_{cov}$) of $50-100\%$. We also choose to adopt both solar and sub-solar metallicity models, as suggested based on the metallicity ranges observed in nearby LIRGs \citep{jrich12}. Our model choices allow us to make consistent comparisons with results from previous studies of YMCs in LIRGs (\citet{stl17,aa20a}) as well as the Legacy Extragalactic UV Survey (LEGUS) of nearby normal star-forming galaxies, which adopt the same {\it yggdrasil} models to determine cluster ages, masses, and extinctions \citep[e.g.,][]{ryon17,mm18a,cook19}.

In the left Panel of Figure 3 we see that the {\it yggdrasil} model tracks decrease vertically from 1 - 5 Myr, approximately perpendicular to the direction of the extinction vector show in the lower-right. The direction and magnitude of this vector are determined using Equation 9 in \citet{salim20}. Following the discussion in Section 3.2.3, we adopt a NIR power-law slope ($\beta$) of -1.6. Studies which sample a variety of MW sightlines find values for $\beta$ ranging from -0.9 to -2.3, with a slight preference to find flatter NIR slopes in regions of higher extinction \citep{fitz09}.

The vertical feature in the SSP model occurs when massive stars evolve from blue supergiants into the Red supergiant (RSGs) evolutionary phase. RSGs are young, massive, luminous stars that rapidly exhaust their core hydrogen. The luminosities of RSGs peak at $\sim 1 \mu$m with absolute J-band magnitudes of $M_{J} = -8$ to $-11$, rivaling the integrated light of Milky Way globular clusters \citep{larsen11}. Thus RSGs, when present, can dominate the NIR flux of a YMC. Therefore the F150W-F200W NIR colors can be used as an absolute age indicator for YMCs younger or older than 5 Myr due to the strong effects RSGs have on the overall continuum shape, and the sharp vertical jump in NIR model colors \citep{gazak13}.

\begin{figure*}[]
\centerline{\includegraphics[width=4.2truein]{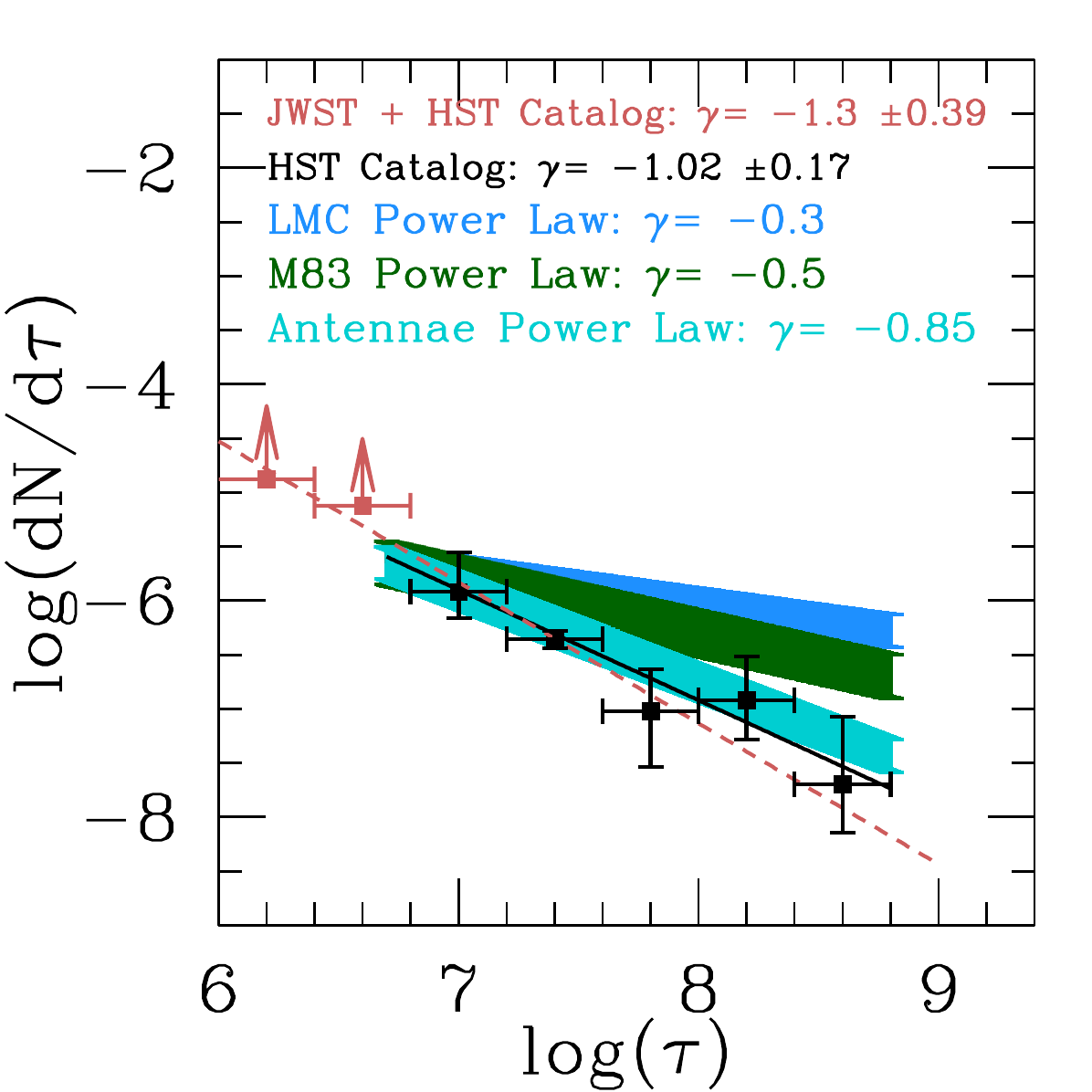}}
\caption{The age distribution function for VV 114 relative to the distribution slopes derived for other nearby galaxies including the LMC (blue), M83 (green), and the Antennae merger (teal) \citep{aab15,mf05}. For clusters with $10^{5.2} < M (M_{\odot}) < 10^{8.0}$ we previously found that the slope of the age distribution from $10-400$ Myr is $\gamma = -1.02 \pm 0.17$ (solid black line). In red we show the age distribution including all YMCs newly-identified in our NIRCam imaging with F150W-F200W $> 0.75$, which corresponds to the color of an un-reddened SSP model at 3 Myr. The age distribution slope which includes these newly-identified clusters is $\gamma = -1.30 \pm 0.39$ from $1 - 10$ Myr. This value is consistent with the uncertainties of the previously determined slope, and confirms that VV 114 displays a steep age gradient even for the youngest clusters identified with {\it JWST}.}
\end{figure*}

\subsection{The Color-Color Diagram}

In the left panel of Figure 3 we show the F150W-F200W vs. F200W-F356W color-color diagram for all 374 compact point-sources identified in our {\it JWST} NIRCam images (grey), as well as the star cluster catalog presented in \citet{linden21} in black. Extended sources, including background galaxies, have been removed from this plot. In purple we show the 135 sources that appear in both catalogs. Overlaid in blue and green are the {\it yggdrasil} SSP models adopted here (with ages marked in red along the sequence) from \citet{yggdrasil}. We note that our UV-optical selection method produced a handful of strongly-detected ($S/N_{F336W} > 5$), young ($t \sim 1 -  3$ Myr), dusty ($A_{V} \sim 3 - 5$), and massive ($M > 10^{5.2} M_{\odot}$) clusters, which have F150W-F200W $> 0.4$ and F200-F356W $\sim -0.7$. This group of YMCs were all subsequently identified with {\it JWST}, and provide an independent verification that our combination of NIR colors can isolate young and dusty sources.

Although the median colors for the two cluster samples are approximately the same, nearly all of the sources that have F200W-F356W $> -0.4$ have been identified in the NIRCam imaging alone. The region marked with a red rectangle in Figure 3 highlights YMC candidates with F150W-F200W $> 0.75$, which corresponds to the color of an un-reddened, solar metallicity, SSP model at 3 Myr. We emphasize that this selection will not identify all compact YMCs within VV 114. Rather, we isolate the 17 reddest NIRCam-detected sources in both F150W-F200W and F200W-F356W colors, which based on existing models strongly suggest they are very young and heavily dust-enshrouded sources. The position and photometry for these 17 sources is given in Table 1. We also find 4 {\it HST}-detected sources with F200W-F356W > 0.2 whose previously-derived cluster ages suggest they are slightly older sources ($t \sim 10$ Myr). We note that these sources are all marginally-detected ($S/N_{F336W} \sim 3$) and therefore have large age/extinction uncertainties.

In the right Panel of Figure 3 we show a zoom-in false-color image of VV 114 marking the locations of 15 of the reddest and newly-discovered YMCs. Notably, these sources lie in the `overlap' region of the merger between the east and west nuclei. Crucially, the extinction vector for $A_{V} = 5$ shown in the lower-right demonstrates that clusters with ages $t = 1 - 5$ Myr will move approximately perpendicular to the models with increasing levels of visual extinction, allowing us to break the age-reddening degeneracy using the combination of NIR colors presented in this analysis. Overall, we stress that this age-dating method is approximate, and a complete SED fit including all available {\it HST} and {\it JWST} observations are left for future analysis.

Based on the extinction vector shown in Figure 3, the reddest YMCs detected with {\it JWST} alone all have visual extinctions of $A_{V}  \geq 15$. However, in Figure 2 we see that diffuse emission surrounds many of the most luminous red sources, and can be found above and below the spiral arm as well as in between individual regions. Including an additional $25 - 50\%$ contribution to the F356W continuum flux density from a 500 K blackbody results in a shift of 0.24-0.44 mag in the measured F200W-F356W colors. Further, an additional contribution to the F356W filter is the polycyclic aromatic hydrocarbon (PAH) feature at 3.3$\mu$m. PAH molecules have been found to be an important component of the interstellar medium, contributing up to $\sim 20\%$ of the total infrared emission from a galaxy \citep{jds07}. Using the averaged {\it Spitzer}+AKARI spectra for star-forming galaxies in the local Universe (1C; \citet{lai20}), we find that the contribution of the 3.3$\mu$m PAH to the F356W filter is $\sim 11 \pm 4\%$. This contribution corresponds to a shift of up to 0.12 mag in the measured F200W-F356W colors (shown with the horizontal vector). Allowing contributions from both hot dust and the 3.3$\mu$m PAH feature, we find that the reddest YMCs in our sample may have extinctions of $A_{V} \sim 5 - 15$. With an unknown relative contribution from both of these components as a function of cluster age or mass we cannot simply use the extinction vector to identify all YMCs candidates. Thus, we adopt the rectangular selection in Figure 3 such that regardless of the exact contributions from hot dust or the 3.3$\mu$m PAH feature, the reddest YMCs selected in both F150W-F200W and F200W-F356W colors that indicate they are very young ($t < 3$ Myr) and dusty ($A_{V} \geq 5$).

The discovery of this `hidden' YMC population quadruples the number of $t < 3$ Myr clusters, and nearly doubles (x1.7) the number of $t < 6$ Myr clusters previously detected with {\it HST} in VV 114. Further, using the {\it yggdrasil} SSP models we take the average $M/L_{F200W}$ ratio from 1 - 3 Myr as well as the Solar AB absolute magnitude at F200W (\citet{willmer18}) to derive stellar masses of $10^{5.8-6.1} M_{\odot}$ for these sources. We note that the M/L ratio varies by less than a factor of 2 over this age range. Further, these masses fall within the range derived from model-fits to the sources in our {\it HST} catalog. However, the possibility remains that all YMCs detected in the `overlap' regions are subject to significant line-of-sight extinction, and therefore have red colors which are unassociated to local dust extinction surrounding each cluster. However, from our {\it HST} catalog we see both young and old clusters with varying levels of internal extinction, as traced by UV-optical emission.

Finally, clusters with F200W-F356W $> -0.4$ and F150W-F200W $\sim 0.4$ are also predominately identified with {\it JWST} NIRCam, and may represent a more evolved, yet still dust-enshrouded population of YMCs in VV 114. However, degeneracy between model ages and colors in this region of the parameter space prevents us from drawing robust conclusions with existing NIRCam observations, because we are unable to determine contamination fraction from sources at older ages. Deriving the timescales necessary for clusters to fully clear their natal gas and dust in local LIRGs will be addressed in a follow-up paper including data from all ERS targets in PID 1328.

\section{Discussion}

The age distribution of stellar clusters in galaxies is the combined result of the cluster formation history and cluster disruption. In a galaxy with a constant cluster formation rate, the slope of the age distribution ($\gamma$) then becomes a measure of how long clusters survive, or at what rate they are being destroyed. If we consider YMCs at a single age, our measured slope implies that their numbers will decrease with time as $dN/dT = - N/\gamma$, where a steeper value for $\gamma$ corresponds to a more rapid cluster destruction.

Studies of the age distribution slope for clusters in nearby galaxies tend to focus on the most massive clusters at ages $10 < \tau (\mathrm{yr}) < 300$ Myr, which are the easiest sources to observe at UV-optical wavelengths \citep{krumholz19}. Therefore, comparisons between different galaxies have been restricted to a limited region of the full cluster age and mass distributions thus far.

In Figure 4 we show the cluster age distribution for VV 114, which contains 90/179 ($\sim 50\%$) confirmed clusters detected at $\geq 3\sigma$ in {\it HST}/F336W, {\it HST}/F435W, and {\it HST}/F184W (black points) with masses $10^{5.2} < M (M_{\odot}) < 10^{8}$. This cut represents the mass-completeness limit for star clusters with ages $t\leq 10^{8.5}$ yr determined from the {\it HST} catalog using artificial source injection and recovery tests for clusters drawn from a distribution of absolute magnitudes \citep{linden21}. This limit is adopted specifically to study the cluster age distribution over a large range in cluster ages, due to the fact that clusters dim as they age, and therefore we increase the lower-mass limit as a function of increasing age in order to retain a mass-complete cluster sample. Thus, our choice of a very conservative mass limit (sources with masses of $10^{4} M_{\odot}$ are detected in both {\it HST} and {\it JWST} imaging) allow us to be confident that the measured slope is not affected by issues related to completeness. We previously derived an age distribution slope from $10 - 400$ Myr of $\gamma = -1.02 \pm 0.17$ (solid black line) with little evidence for significant deviations as a function of age. We note that this value is steeper than the slopes observed in both LMC (blue) and M83 (green) \citep{aab15}, as well as observations of the Antennae merger galaxy (teal) \citep{mf05}, providing evidence that the slope of the age distribution tends to be steep in starburst and actively-merging galaxies in the local Universe.

Using our {\it HST}-derived cluster catalog we found 4 clusters with $t < 2.5$ Myr and 26 with $2.5 - 6$ Myr. The selection of YMC candidates in Figure 3 is meant to isolate sources whose F150W-F200W and F200W-F356W colors suggest they must be young, massive, and heavily dust-enshrouded sources. In red we show the updated age distribution including all 17 YMCs with F150W-F200W $> 0.75$ identified in our NIRCam imaging (corresponding to an SSP model at 3 Myr) assigned to the youngest age-bin. We additionally include the 4 sources (purple points in selection box with F200W - F356W $> 0.2$), whose previously-derived ages with {\it HST} suggested they were slightly older (with large uncertainty), assigned to the second youngest age-bin. If we assume the 17 newly-identified sources all have ages between 1 - 3 Myr, the average stellar mass of these sources is $10^{5.9} M_{\odot}$, which is well-above above our {\it HST}-derived completeness limit of $10^{5.2} M_{\odot}$.

Given the conservative nature of our color selection method we are likely excluding sources with F150W-F200W $< 0.75$, which are also young ($t < 6$ Myr), and therefore the values in Figure 4 for the youngest age-bins represent lower-limits. One additional source of uncertainty is effect blending has on the derived stellar masses of individual YMC candidates. At the resolution of our observations, multiple lower mass clusters may appear as a single massive source. In \citet{stl17} we examined the effect cluster blending has on the number of extracted sources in the Antennae merger \citep[$D_{L} = \sim 24$ Mpc;][]{bcm10} by progressively smoothing {\it HST} F435W, F555W, and F814W images. At the distance of VV 114 ($84.4$Mpc) we find that $\sim 25\%$ of sources with stellar mass $\geq 10^{6} M_{\odot}$ would be identified as a two or more clusters at the distance of the Antennae. However, we note that reducing the stellar mass of all YMCs with F150W-F200W $> 0.75$ by half results in an average of $10^{5.6} M_{\odot}$, which is still above our completeness limit.

For sources which are only detected with NIRCam, and have $0.1 < F150W-F200W < 0.75$, the possible range of ages is much larger (5 - 100 Myr) due to the degeneracy of the SSP models in the NIR and therefore the contamination from sources at older ages. Taking the median age of sources in this cloud to be 10 Myr we derive an average stellar mass of $10^{5.1} M_{\odot}$. We stress that because our completeness limit is sufficiently high, this population of {\it JWST} sources excluded in our analysis, due to degeneracy in the derived ages, will have even lower stellar masses and ultimately should not impact the results presented in Figure 4. The age distribution slope which includes the reddest YMCs is $\gamma = -1.30 \pm 0.39$ for $10^{6} < \tau (\mathrm{yr}) < 10^{7}$ yr. This value is consistent with the previously determined slope within the uncertainties on both measurements, and confirms that VV 114 has a steep age gradient for all massive star clusters across the entire range of cluster ages observed. This result provides further evidence that steep age distribution slopes are found predominately in starburst and merging galaxies.

We note that the steep slope observed in VV 114 may actually be due to an increase in YMC formation as opposed to cluster destruction over the last $10^{8.5}$yr. However, CALIFA IFU observations of VV 114 suggest that the extinction-corrected enhancement of the star formation rate over the last 300 Myr relative to normal star-forming galaxies is a only factor of $\sim 2 - 3$ in the central half-light radius \citep{cf17}, which would result in a change in $\gamma$ of $\sim -0.2$. We therefore attribute the much steeper slope observed in VV 114 to rapid cluster destruction as opposed to changes in the YMC formation rate.

Finally, the lack of significant deviations in the shape of the cluster age distribution suggests that the cluster disruption rate in VV 114 has not changed dramatically over the last 0.5 Gyr. For YMCs with ages $t < 5$ Myr, supernova will not have had a sufficient amount of time to unbind the cluster, whereas feedback from clusters with ages $t>10$ Myr will be dominated by supernova. In Figure 4 we see that the age distribution slope in these two regimes is similar, indicating that the important feedback channel for cluster disruption is the early winds/radiation pressure at $t< 5$ Myr, and/or the overall galactic environment sets the destruction rate (as parameterized by $\gamma$) throughout the merger. Future {\it JWST} observations of YMCs in LIRGs, which span a range in SFR, merger stages, and distance from the SFMS, will be crucial to uncovering what truly sets the balance between YMC formation and evolution in extreme star-forming galaxies.

\section{Summary}

We have presented the results of a {\it James Webb Space Telescope} NIRCam investigation into the young massive star cluster population in the luminous infrared galaxy VV 114. We identify 374 compact YMC candidates with $FWHM_{F150W} -   FWHM_{PSF} < 2\sigma_{FWHM,maj}$ as well as a $S/N \geq 3$, 5, and 5 at F150W, F200W, and F356W respectively for which we perform aperture photometry. Based on the derived NIR magnitudes and colors of our new YMC catalog we reach the following conclusions:

\noindent
(1). We find that $20\%$ of the {\it JWST}-identified star clusters are undetected at optical wavelengths. This fraction increases to $60\%$ at $\lambda=0.336\mu$m, and demonstrates that although UV observations are necessary for {\it HST}-based studies of star clusters in LIRGs, this approach does miss large fractions of highly-embedded YMCs.

\noindent
(2). Based on {\it yggdrasil} SSP models we newly-identify 17 YMCs in our {\it JWST} catalog with F150W-F200W and F200W-F356W colors suggesting they are all very young, dusty ($A_{V} \sim 5 - 15$), and massive ($10^{5.8} < M (M_{\odot}) < 10^{6.1}$). The inclusion of these `hidden' sources quadruples the number of $t < 3$ Myr clusters, and nearly doubles (x1.7) the number of $t < 6$ Myr clusters previously detected with {\it HST} in VV 114. Finally, we identify a group of NIRCam-detected clusters with F200W-F356W $> -0.4$ and F150W-F200W $\sim 0.4$, which may represent a more evolved, yet still dust-enshrouded population of YMCs in VV 114.

\noindent
(3). The resulting cluster age distribution slope of $dN/d\tau \propto \tau^{-1.30 +/- 0.39}$ for $10^{6} < \tau (\mathrm{yr}) < 10^{7}$ is consistent with the previously determined value of $\gamma = -1.02 \pm 0.17$ for $10^{7} < \tau (\mathrm{yr}) < 10^{8.5}$, and confirms that VV 114 has a steep age gradient for all massive star clusters across the entire range of cluster ages observed. The consistency between our {\it JWST}- and {\it HST}-derived age distribution slopes may imply that feedback from stellar winds and supernovae are not able to significantly impact the overall balance of formation and destruction of YMCs in VV 114 over the last 0.5 Gyr.

\begin{acknowledgements}

All of the data presented in this paper were obtained from the Mikulski Archive for Space Telescopes (MAST) at the Space Telescope Science Institute. S.T.L was partially supported thorough NASA grant HST-GO16914. A.S.E. was supported by NSF grant AST 1816838 and by NASA through grants HST-GO10592.01-A, HST-GO11196.01-A, and HST-GO13364 from the Space Telescope Science Institute, which is operated by the Association of Universities for Research in Astronomy, Inc., under NASA contract NAS5- 26555. V.U. acknowledges funding support from NASA Astrophysics Data Analysis Program (ADAP) grant 80NSSC20K0450. H.I. and T.B. acknowledge support from JSPS KAKENHI Grant Number JP21H01129 and the Ito Foundation for Promotion of Science.~F. K. acknowledges support from the Spanish program Unidad de Excelencia Maria de Maeztu CEX2020-001058-M, financed by MCIN/AEI/10.13039/501100011033. This work was also partly supported by the Spanish program Unidad de Excelencia Mari\'a de Maeztu CEX2020-001058-M, financed by MCIN/AEI/10.13039/501100011033. This research has also made use of the NASA/IPAC Extragalactic Database (NED) which is operated by the Jet Propulsion Laboratory, California Institute of Technology, under contract with the National Aeronautics and Space Administration. Finally, the National Radio Astronomy Observatory is a facility of the National Science Foundation operated under cooperative agreement by Associated Universities, Inc.

\end{acknowledgements}

\bibliography{master_ref}{}
\bibliographystyle{aasjournal}



\begin{deluxetable*}{lllllllll}
\center
\tablecaption{NIRCam-Detected Red YMC Catalog \label{tbl-1}}
\tabletypesize{\footnotesize}
\tablewidth{0pt}
\tablehead{
\colhead{ID} & \colhead{RA (J2000)}  & \colhead{Dec (J2000)} & \colhead{F150W} & \colhead{$\sigma_{F150W}$}  & \colhead{F200W} & \colhead{$\sigma_{F200W}$} & \colhead{F356W} &  \colhead{$\sigma_{F356W}$}}
\startdata
1 & 16.945622 & -17.507619 & 21.54 & 0.05 & 20.80 & 0.06 & 20.78 & 0.09 \\
2 & 16.945276 & -17.507493 & 21.09 & 0.04 & 20.34 & 0.03 & 20.92 & 0.10 \\
3 & 16.946065 & -17.507593 & 20.66 & 0.06 & 19.99 & 0.08 & 20.90 & 0.09 \\
4 & 16.946339 & -17.507373 & 22.27 & 0.07 & 20.99 & 0.05 & 21.13 & 0.06 \\
5 & 16.946136 & -17.507442 & 21.41 & 0.09 & 20.01 & 0.06 & 19.66 & 0.02 \\
6 & 16.946707 & -17.507150 & 21.58 & 0.04 & 20.48 & 0.04 & 20.66 & 0.07 \\
7 & 16.942848 & -17.507312 & 21.74 & 0.07 & 20.68 & 0.05 & 20.66 & 0.06 \\
8 & 16.947392 & -17.507028 & 21.35 & 0.07 & 20.39 & 0.06 & 20.57 & 0.03 \\
9 & 16.947433 & -17.507265 & 20.56 & 0.15 & 19.17 & 0.05 & 18.70 & 0.04 \\
10 & 16.946946 & -17.507304 & 21.36 & 0.06 & 20.42 & 0.02 & 19.71 & 0.05 \\
11 & 16.947055 & -17.507350 & 20.34 & 0.06 & 19.70 & 0.09 & 19.77 & 0.14 \\
12 & 16.946361 & -17.506840 & 21.49 & 0.08 & 20.18 & 0.06 & 19.32 & 0.03 \\
13 & 16.946557 & -17.507075 & 22.11 & 0.05 & 20.90 & 0.04 & 20.89 & 0.07 \\
14 & 16.946062 & -17.506642 & 21.28 & 0.08 & 20.33 & 0.07 & 20.54 & 0.06 \\
15 & 16.946195 & -17.506178 & 21.72 & 0.03 & 20.45 & 0.02 & 20.70 & 0.03 \\
16 & 16.946779 & -17.506351 & 21.12 & 0.03 & 20.12 & 0.02 & 19.78 & 0.03 \\
17 & 16.943581 & -17.508801 & 22.17 & 0.09 & 21.56 & 0.07 & 21.40 & 0.03 \\
\enddata
\tablecomments{All apparent magnitudes are given in the AB photometric system}
\end{deluxetable*}

\end{document}